\documentclass[a4paper]{aa}
\usepackage{psfig,times}
\usepackage{amssymb}
\usepackage{graphicx}
\usepackage{txfonts}
\begin{document}
\title{Magnetic shear-driven instability and turbulent mixing in magnetized 
       protostellar disks} 
\author{A.~Bonanno\inst{1,2}, V.~Urpin\inst{1,3}}
\institute{$^{1)}$ INAF, Osservatorio Astrofisico di Catania,
           Via S.Sofia 78, 95123 Catania, Italy \\
           $^{2)}$ INFN, Sezione di Catania, Via S.Sofia 72,
           95123 Catania, Italy \\
           $^{3)}$ A.F.Ioffe Institute of Physics and Technology and
           Isaac Newton Institute of Chile, Branch in St. Petersburg,
           194021 St. Petersburg, Russia}

\date{\today}

\abstract
{Observations of protostellar disks indicate the presence of the magnetic 
field of thermal (or superthermal) strength. In such a strong magnetic 
field, many MHD instabilities responsible for turbulent transport of the 
angular momentum are suppressed.} 
{We consider the shear-driven instability 
that can occur in protostellar disks even if the field is superthermal.}
{This instability is caused by the combined influence of shear and 
compressibility in a magnetized gas and can be an efficient mechanism  
to generate turbulence in disks.}
{The typical growth time is of the order of several rotation periods.}
{}  


\authorrunning{A. Bonanno \& V. Urpin}
\titlerunning{Shear-driven instability in protostellar disks}

\maketitle

\section{Introduction}

Protostellar disks require sufficiently strong turbulence to enhance the 
efficiency of angular momentum transport. The origin of turbulence is 
often attributed  to hydrodynamic and hydromagnetic instabilities that can 
arise in differentially-rotating, stratified gaseous disks. One of the 
candidates is  magnetorotational instability (MRI), which  can operate in 
a conductive flow if the angular velocity decreases with the cylindrical 
radius and the magnetic field is not strong (Velikhov 1959). The MRI has 
been studied in depth in the case of  stellar  
and accretion disk conditions (see, 
e.g., Fricke 1969; Safronov 1969; Acheson 1978, 1979; Balbus \& Hawley 1991; 
Kaisig et al. 1992; Kumar et al. 1994; Zhang et al. 1994). Simulations of 
the MRI in disks (Hawley et al. 1995, Matsumoto \& Tajima 1995, Brandenburg 
et al. 1995, Torkelsson et al. 1996, Arlt \& R\"udiger 2001) show  the 
turbulence generated can significantly enhance  the angular momentum transport. 

Most likely, however, the number of instabilities that can arise in 
astrophysical disks is quite large. An analysis of MHD modes in stratified 
accretion disks demonstrates a wide variety of instabilities even in the 
case of simple magnetic geometry (Keppens, Casse \& Goedbloed 2002). 
Therefore, the current point of view on the origin of turbulence in disks 
is likely highly simplified. Even pure, hydrodynamic origin of turbulence 
cannot be excluded (see Urpin 2003, Arlt \& Urpin 2004, Dubrulle et 
al. 2005) despite the most efficient local linear hydrodynamic instabilities 
advocated to date are not sufficiently efficient. Global instabilities, such 
as the baroclinic-like instability of Klahr \& Bodenheimer (2003), are 
sensitive to boundary conditions (Johnson and Gammie 2006) and, therefore, 
are unlikely to drive turbulence in disks. { Note, however, that
such factors as fast radiative cooling, high thermal diffusion, and large
radial temperature gradients can strengthen the baroclinic feedback and
make the baroclinic instability more efficient (Petersen et al. 2007).}
Astrophysical disks are stratified 
and stratification can change stability properties of shear 
flows, providing either a stabilizing or destabilizing effect, depending on
details of the disk structure. Recently, Dubrulle 
et al. (2005) have considered one of the possible ``strato-rotational'' 
instabilities arising in the presence of both differential rotation and stable 
stratification. However, Brandenburg \& R\"udiger (2005) demonstrated that 
the growth rate of this instability decreases with an increasing Reynolds number, 
rendering the instability less relevant for astrophysical applications. 
Note also that convection can drive turbulence in disks with 
unstable stratification, but it induces inward turbulent transport of the 
angular momentum instead of the required outward one (see, e.g., Stone \& 
Balbus 1996). 

As it has been argued by Lesur \& Longaretti (2005),
nonlinear hydrodynamic instability that often occurs in linearly-stable 
flows at sufficiently large Reynolds numbers is likely inefficient in 
disks even for the most optimistic extrapolations of numerical data. 
At least, the subcritical transition to 
hydrodynamic turbulence cannot occur in quasi-keplerian flows at Reynolds 
numbers up to $\sim 10^6$ (Ji et al. 2006). Of course, Reynolds numbers are much 
higher in real disks; nevertheless it seems that there is little room for 
nonlinear instability.    

The stability properties of protostellar disks are much different from those of 
accretion disks. The magnetic Reynolds number is likely not very large in 
cold- and dense-protostellar disks because of a low electrical conductivity, 
and so the field cannot be treated as "frozen" into the gas (Gammie 1996). The 
effect of Ohmic dissipation on the MRI has been considered in the linear 
(Jin 1996) and nonlinear regimes (Sano, Inutsuka \& Miyama 1998; Drecker, 
R\"udiger \& Hollerbach 2000). Fleming \& Stone (2003) and Turner et al. (2006) 
treated turbulent mixing caused by the MRI in protostellar disks,  taking into 
account magnetic diffusivity. They found that the midplane is shielded from 
cosmic rays, and that MRI does not occur even under the most favorable 
conditions. Nevertheless, turbulence can mix a fraction of the weakly-conducting 
surface material into the interior, providing some coupling of the 
midplane gas to the magnetic field. As a result, weak stresses can appear in 
the disk midplane.
  
As first pointed out by Wardle (1999), poorly conducting protostellar 
disks can be strongly magnetized if electrons are the main charge carriers. 
Therefore, transport must be  anisotropic with substantially different 
properties along and across the magnetic field. 
In strongly magnetized gas the Hall component provides the main contribution
to the resistivity tensor, which produces  the electric field perpendicular to both
the magnetic and electric current.
The stability analysis by Wardle  (1999) shows 
that the Hall effect can provide either stabilizing or destabilizing 
influences depending on the direction of the field. 
%
Balbus \& Terquem (2001)  have conducted a more general study of the role of the Hall term
on MRI.  They found that the Hall effect qualitatively changes the stability properties
of protostellar disks and can lead to instability even if the angular 
velocity increases outward.  These authors, however, did not take into account 
the effect of gravity that is crucial for disks. 
Urpin \& R\"udiger (2005) considered MRI under the combined influence of the Hall effect and gravity, and
also derived also the criteria of several other instabilities that can occur in protostellar disks. 
Sano and Stone (2002) investigated the effect of the Hall term on the evolution of the MRI in weakly-ionized disks 
using local axisymmetric simulations. 
These authors concluded  the properties of the  MRI depend essentially on 
the direction of the magnetic field as it is anticipated from the dispersion 
equation in a linear stability analysis. 
Salmeron \& Wardle (2005) have also considered the properties of the MRI modified by the Hall effect.  
These authors argued that the MRI is active in protoplanetary disks over a 
wide range of field strengths and fluid conditions. The Hall conductivity 
results in a faster growth of perturbations and extends the region of 
instability. Recently, Livertz et al. (2007) and Shtemler et al. (2007) have 
considered the Hall MRI in a non-axisymmetric case. This type of instability 
is proposed as a viable mechanism for the azimuthal fragmentation  of the 
protoplanetary disks and planet formation. The non-axisymmetric instability 
is caused  by the combined effect of the radial stratification and 
Hall electric field. Note that the MRI and its modifications can be completely 
suppressed if the magnetic field is sufficiently strong (see Urpin 1996, 
Kitchatinov \& R\"{u}diger 1997). 

If rotation is cylindrical and $\Omega = \Omega(s)$, where $\Omega$ is the 
angular velocity and $s$ is the cylindrical radius, the critical magnetic 
field that suppresses the MRI is given approximately by the condition 
$c_{A} / H \sim s \Omega' \sim \Omega$ where $c_{A} = B/ \sqrt{ 4 \pi \rho}$
is the Alfv\'en velocity. Since $\Omega \sim c_s /H$ in the standard disk,  
where $c_s$ is the sound speed, we find that the MRI is suppressed if the 
magnetic field is superthermal, $c_{A} > c_s$. Recent measurements
(Hutawarakorn \& Cohen 1999, 2005, Donati et al. 2005) indicate that the 
magnetic field can be strong in protostellar disks. This particularly
concerns the innermost regions where the field strength reaches $\sim 1$ kG 
at the radius $\sim 0.05$ AU (Donati et al. 2005). The field configuration 
includes the azimuthal component whose direction agrees with  
the radial field  sheared by the disk differential rotation. The 
derived ratio of the azimuthal and radial fields in the surface layers is 
$\sim 0.5$. Note that the azimuthal component can be substantially stronger 
than the radial one in the disk interiors because dynamo theories 
(such as the $\alpha - \Omega$ dynamo) usually predict that the generated 
field should have a strong toroidal component. According to estimates by 
Donati et al. (2005), the equipartition field strength with roughly equal 
thermal and magnetic pressures should be $\sim 10^3$ Gauss at the radius $\sim 0.05$ 
AU in protostellar disks. This indicates the detected field is approximately of 
the thermal or even superthermal strength and can substantially 
modify or even suppress the MRI.

Recently, Bonanno \& Urpin (2006, 2007) have argued that a compressible, 
differentially-rotating flow is unstable if the magnetic field has a
non-vanishing, radial component. A remarkable feature of this shear-driven
instability is that it can arise even if the magnetic field is superthermal. 
Bonanno \& Urpin (2006, 2007) have considered the instability in the case of 
a highly conductive plasma when the magnetic diffusivity plays no role.
However, the instability can also arise in weakly-ionized protostellar
disks where the magnetic field has a radial component detected in observations
(Donati et al. (2005). The effect of a finite diffusivity can be important for
the onset of instability and for the properties of generated turbulence,
particularly near the the mid-plane where conductivity is extremely low. In 
this paper, we consider the shear-driven instability in the conditions of 
protostellar disks and show that the conditions of instability in weakly and
highly ionized plasmas differ substantially. Nevertheless, the shear-driven
instability can manifest itself even in weakly-ionized protostellar disks. 

The paper is organized as follows. 
In Section 2, we consider the basic equations governing instability in 
compressible dissipative fluids and we then derive the dispersion relation. In 
Section 3, we derive the criteria of instability and discuss  conditions 
under which instability can occur in protostellar disks. The growth rate  
of instability is calculated in Section 4. Finally, in Section 5, we discuss
the results obtained.

\section{Basic equations and dispersion relation}
The electrical conductivity is low in protostellar disks- and 
the magnetic field cannot be considered as ``frozen'' into the gas. The 
magnetic diffusivity is given by $\eta =c^{2} m_{e}/ 4 \pi e^{2} n_{e} \tau$ 
where $m_{\rm e}$ and $n_{\rm e}$ are the mass and number density of 
electrons, respectively, and $\tau$ is their relaxation time (see, e.g., 
Spitzer 1978). In protostellar disks, $\tau$ is determined by the scattering 
of electrons on neutrals, then $\tau = 1/ n \langle \sigma v \rangle$ where 
$\langle \sigma v \rangle$ is the average product of the cross-section and 
velocity, and $n$ is the number density of neutrals. Using the fitting 
expression for $\langle \sigma v \rangle$ obtained by Draine et al. (1983), 
we have
\begin{equation}
\eta = 2.34 \times 10^{3} f^{-1} T_{2}^{1/2} \;\; {\rm cm}^{2} \;
{\rm s}^{-1},
\end{equation} 
where $f=n_{\rm e}/n$ is the ionization fraction, and $T_{2} = T/100$ K, with 
$T$ being the temperature. 

The magnetic diffusivity can be anisotropic in some regions of protostellar 
disks (Wardle 1999) because the electron gas is magnetized despite a low 
temperature. The effect of the magnetic field on transport properties is 
usually characterized by the magnetization parameter $a_{\rm e}(B)= \omega_B 
\tau$, where $\omega_B= e B /m_{\rm e} c$ is the gyrofrequency of electrons 
(see, e.g., Spitzer 1978). Again, using the fitting formula by Draine et al. 
(1983), we can estimate the magnetization parameter as
\begin{equation}
a_{\rm e} (B) \approx 21\ B n_{14}^{-1} T_{2}^{-1/2},
\end{equation} 
where the magnetic field $B$ is measured in Gauss and $n_{14}= n/10^{14}$ 
cm$^{-3}$. If $a_{\rm e}(B) > 1$, i.e. 
\begin{equation}
B>  0.048 \ n_{14} \ \sqrt{T_2} \;\; {\rm Gauss},
\end{equation} 
then the electron transport is anisotropic and the magnetic diffusivity is 
represented by a tensor. In plasma of protostellar disks, the difference 
between components of the magnetic diffusivity parallel and perpendicular 
to the magnetic field  is small (see, e.g., Balbus \& Terquem 2001). 
The Hall component is described as
\begin{equation}
a_{\rm e}(B) \eta = {cB \over 4\pi e n_{\rm e}}.
\end{equation}  

We work in cylindrical coordinates ($s$, $\varphi$, $z$) with the 
unit vectors ($\vec{e}_{s}$, $\vec{e}_{\varphi}$, $\vec{e}_{z}$).
The equations of compressible MHD read:  
\begin{eqnarray}
\dot{\vec{v}} + (\vec{v} \cdot \nabla) \vec{v} = - \frac{\nabla p}{\rho} 
+ \vec{g}  + \frac{1}{4 \pi \rho} (\nabla \times \vec{B}) \times \vec{B}, 
\end{eqnarray}
\begin{equation}
\dot{\rho} + \nabla \cdot (\rho \vec{v}) = 0, 
\end{equation}
\begin{equation}
\dot{p} + \vec{v} \cdot \nabla p + \gamma p \nabla \cdot 
\vec{v} = 0,
\end{equation}
\begin{eqnarray}
\dot{\vec{B}} - \nabla \times (\vec{v} \times \vec{B}) 
+ \frac{c}{4 \pi e} \nabla \times \left[
\frac{1}{n_{\rm e}} (\nabla \times \vec{B}) \times \vec{B} \right] \nonumber \\
+ \eta \nabla \times (\nabla \times \vec{B}) = 0,
\end{eqnarray}
\begin{equation}
\nabla \cdot \vec{B} = 0. 
\end{equation} 
Our notation is as follows: $\rho$ and $\vec{v}$ are  gas density and 
velocity, respectively; $p$ is pressure;  $\gamma$ is the adiabatic 
index; and $\vec{g}$ is gravity. The third term on the l.h.s. of Eq.~(8)
describes the Hall effect. We cannot neglect this term 
because the magnetization parameter $a_e$ is large in protostellar disks.
However, in what follows, we will consider a special type of perturbation
for which the Hall effect is unimportant.  

The basic state on which the stability analysis is performed is assumed to 
be quasi-stationary with the angular velocity $\Omega = \Omega(s)$ and 
$\vec{B} \neq 0$. We assume  hydrostatic equilibrium for  
basic state:
\begin{equation}
\frac{\nabla p}{\rho} = \vec{D} + \frac{1}{4 \pi \rho} 
(\nabla \times \vec{B}) \times \vec{B} \;\;, \;\;\;\;
\vec{D} = \vec{g} + \Omega^{2} \vec{s}.
\end{equation}
We consider magnetic configurations where both the radial and azimuthal 
field components are present. The presence of a radial magnetic field 
and differential rotation leads to the development of the azimuthal field. 
In protoplanetary disks, however, the rate of stretching of the azimuthal 
field from $B_s$ is reduced by ohmic dissipation and Hall effect. 
Then, Eq.~(8) yields the following steady-state condition:
\begin{eqnarray}
[\nabla \times (\vec{v} \times \vec{B})]_{\varphi} =
\! \left\{  \! \nabla \!  \times \!\left[ \! 
\frac{c}{4 \pi e n_{\rm e}} \! 
(\! \nabla \! \times \! \vec{B}) \! \times \! \vec{B} \! \right]  \! 
\right\}_{\varphi}  
\nonumber \\
+ \eta [\nabla \times (\nabla \times \vec{B})]_{\varphi}.
\end{eqnarray}
Assuming that $B_{\varphi} > B_{s}$, Eq.~(11) can be transformed into:
\begin{eqnarray}
\frac{c}{4 \pi e} \left\{ \frac{\partial}{\partial s} \left( 
\frac{B_{\varphi}}{n_{\rm e}} \frac{\partial B_{\varphi}}{\partial z} \right)
-\frac{\partial}{\partial z} \left[ \frac{B_{\varphi}}{n_{\rm e}}
\left( \frac{\partial B_{\varphi}}{\partial s} + \frac{B_{\varphi}}{s}
\right) \right] \right\}
\nonumber \\ 
- \eta \left( \Delta - \frac{1}{s^2} \right) B_{\varphi} = s \Omega' B_{s}.
\end{eqnarray}
If the Hall effect is negligible, the first term on the l.h.s. of this
equation is small, and the generated toroidal field is stronger than 
the radial field by a factor of the order of the magnetic Reynolds number, 
Re$_m$, 
\begin{equation}
B_{\varphi} \sim {\rm Re}_m B_s,
\end{equation}
where Re$_m = s \Omega' H^2/ \eta \sim \Omega H^2/\eta$ with $H$ being the 
half-thickness of the disk (we assume $s \Omega' \sim \Omega$). If the 
magnetic field is strong and the Hall effect dominates ohmic dissipation, 
then one estimates from Eq.~(12) that,
\begin{equation}
B_{\varphi} \sim \frac{{\rm Re}_m^{1/2}}{a_{\rm e}^{1/2}(B_s)} \left( 
\frac{s}{H} \right)^{1/2} B_s.
\end{equation}
In strongly magnetized disks, the generated toroidal field can be substantially
weaker than what follows from the simplest estimate (13). Generally, even if 
Re$_m \gg 1$, the toroidal field is comparable to the radial one if
$a_{\rm e} \sim {\rm Re}_m$ despite a strong differential rotation. It is 
possible, for example, that the toroidal and poloidal fields measured in FU 
Orionis are comparable because of the Hall effect.  

We consider the stability of axisymmetric short-wavelength perturbations.
Small perturbations will be indicated by subscript 1, while unperturbed 
quantities will have no subscript. The linearized momentum equation is:
\begin{eqnarray}
\dot{\vec{v}}_{1} + (\vec{v}_{1} \cdot \nabla) \vec{v} + (\vec{v} \cdot
\nabla) \vec{v}_{1} = - \frac{\nabla p_{1}}{\rho} + \frac{\rho_1}{\rho}
\left[ \frac{\nabla p}{\rho} - \right.
\nonumber \\ 
\left. \frac{1}{4 \pi \rho} (\nabla \times \vec{B}) \times \vec{B} \right] 
+ \frac{1}{4 \pi \rho} (\nabla \times \vec{B}_{1}) \times \vec{B} 
\nonumber \\
+ \frac{1}{4 \pi \rho} (\nabla \times \vec{B}) \times \vec{B}_{1}. 
\end{eqnarray}
Taking into account that the unperturbed motion is rotation ($\vec{v} 
= s \Omega \vec{e}_{\varphi}$) and using Eq.~(10), we can transform this 
equation into 
\begin{eqnarray}
\dot{\vec{v}}_{1} + 2 \Omega \times \vec{v}_{1} + \vec{e}_{\varphi} s \Omega'
v_{1s} = - \frac{\nabla p_{1}}{\rho} + \frac{\rho_{1}}{\rho} \vec{D} 
\nonumber \\
+ \frac{1}{4 \pi \rho} [(\nabla \times \vec{B}_{1}) \times \vec{B} 
+ (\nabla \times \vec{B}) \times \vec{B}_{1}]. 
\end{eqnarray}
The term proportional to $\rho_1/\rho$ is typically small in a short-wavelength
approximation. Indeed, we can estimate $p_1 \sim c_s^2 \rho_1$ where $c_s$
is the sound speed. Then, the pressure term in Eq.~(16) is $(c_s^2/ \lambda)
(\rho_1/\rho)$ where $\lambda$ is the lengthscale of perturbations. 
Observations indicate that the angular velocity can be smaller than
the Keplerian angular velocity at least in a fraction of the disk 
volume (Donati et al. 2005) and, hence, $D \sim g$. Note that even departures
by a factor of 2-3 from the Keplerian velocity reported by Donati et al. (2005)
can produce a substantial increase in the local disk thickness compared to 
the ``canonical'' value $\sim 0.1 s$. Comparing the first two terms on the r.h.s. 
of Eq.~(16), we conclude that the pressure term is larger if $c_s^2 / \lambda 
\gg D \sim g$, or $\lambda \ll c_s^2/g$. Since $c_s^2/g$ is the pressure 
scaleheight, this inequality is equivalent to the condition of applicability 
of a short wavelength approximation. Therefore, we can neglect the term 
proportional to $\vec{D}$ in what follows. An additional simplification is related 
to the last term on the r.h.s. of Eq.~(16) (the ``curvature term'' according 
to Pessah \& Psaltis 2005). The toroidal field can often be stronger than 
the poloidal one and, as a result, the influence of unperturbed electric 
currents can be important. However, if the toroidal field satisfies the 
condition:
\begin{equation}
B_{\varphi} < (s/ \lambda) \max(B_s, B_z),
\end{equation}   
the contribution of ``the curvature term'' is small and can be neglected.

Following similar transformations with the remaining Eqs.~(6)-(9), we 
arrived at the linearized equations needed for stability analysis. We consider 
perturbations with the space-time 
dependence $\propto \exp ( \sigma t - i \vec{k} \cdot \vec{r})$, where 
$\vec{k}= (k_{s}, 0, k_{z})$ is the wavevector. Then, the linearized 
MHD-equations read with accuracy in the lowest order in $\lambda/s$:  
\begin{eqnarray}
\sigma \vec{v}_{1} + 2 \vec{\Omega} \times \vec{v}_{1}
+ \vec{e}_{\varphi} s \Omega' v_{1s}    
= \frac{i \vec{k} p_{1}}{\rho}  
\nonumber \\
- \frac{i}{4 \pi \rho} (\vec{k} \times \vec{B}_{1}) \times \vec{B} ,  
\end{eqnarray}
\begin{equation}
\sigma \rho_{1} - i \rho (\vec{k} \cdot \vec{v}_{1}) = 0, 
\end{equation}
\begin{equation}
\sigma p_{1} -i \gamma p (\vec{k} \cdot \vec{v}_{1}) = 0, 
\end{equation}
\begin{eqnarray}
(\sigma + \omega_{\eta}) \vec{B}_{1} = \vec{e}_{\varphi} s \Omega' 
B_{1s } -i (\vec{B} \cdot \vec{k}) \vec{v}_{1} + i \vec{B} (\vec{k} 
\cdot \vec{v}_{1})
\nonumber \\
- \frac{c (\vec{k} \cdot \vec{B})}{4 \pi e n_{\rm e}} \vec{k} \times \vec{B}, 
\end{eqnarray}
\begin{equation}
\vec{k} \cdot \vec{B}_{1} = 0,
\end{equation}
where $\omega_{\eta}= \eta k^2$.

The dispersion equation, corresponding to Eqs.~(18)-(22) is rather complex in 
the general case. Therefore, we consider only a particular case when the 
wavevector is perpendicular to the magnetic field, $\vec{k} \cdot \vec{B} = 
0$. After some algebra, Eqs.~(18)-(22) can be combined into a sixth-order 
dispersion relation,
\begin{equation}
\sigma^{6} + a_{5} \sigma^5 + a_{4} \sigma^{4} + a_{3} \sigma^{3} + 
a_{2} \sigma^{2} + a_{1} \sigma + a_{0}= 0. 
\end{equation}
The coefficients of this equation are expressed in terms of 
characteristic frequencies,
\begin{eqnarray}
a_{5}= 2 \omega_{\eta}, \;\; a_{4} = \omega_{\eta}^2 + \omega^{2}_{0} + 
\kappa^{2}, \nonumber \\
a_{3} = \omega^{3}_{B \Omega} + \omega_{\eta} (\omega_0^2 +
\omega_s^2 + 2 \kappa^2), \nonumber \\
a_{2} = \omega^{2}_{\eta} (\omega^{2}_{s} + \kappa^{2}) +
\mu \kappa^{2} \omega^{2}_{0}, \nonumber \\
a_{1} = \mu \kappa^2 [ \omega^{3}_{B \Omega} + \omega_{\eta} (
\omega^{2}_{s} + \omega^2_{0})], \;\;
a_{0} = \mu \omega^{2}_{s} \omega^{2}_{\eta} \kappa^{2},  \nonumber
\end{eqnarray}
where $\mu = k^{2}_{z}/k^{2}$. The characteristic frequences are given by
\begin{eqnarray}
\kappa^{2} = 4 \Omega^{2} + 2 s \Omega \Omega' \;, \;\; 
\omega^{2}_{0}= \omega^{2}_{s} + \omega^{2}_{m} \;, \;\; \omega^{2}_{s} = 
c^{2}_{s} k^{2} , \nonumber \\
\omega^{2}_{m} = c^{2}_{A} 
k^{2} \;, \;\;\; \vec{c}_{A} =
\vec{B} /\sqrt{4 \pi \rho} \;, \;\;
\omega^{3}_{B \Omega} = k^{2} c_{A \varphi} c_{A s} s \Omega' 
\nonumber,
\end{eqnarray}
where $c_{s} = \sqrt{\gamma p/ \rho}$ is the sound speed. Eq.~(23) 
describes six modes that can generally exist in a compressible,
rotating, magnetized gas. 

In the non-dissipative, incompressible, limit when the sound speed $c_{s}$ is
very large, Eq.~(23) reduces to the dispersion relation for the inertial 
waves, 
\begin{equation}
\sigma^{2} (\sigma^{2} + \mu \kappa^{2}) =0.
\end{equation}
This equation allows for unstable solutions if the Rayleigh criterion is
fulfilled, $\kappa^{2} < 0$. Note that Eq.~(23) does not describe the 
magnetorotational instability since it does not occur for perturbations with 
$\vec{k} \cdot \vec{B} = 0$.

In non-dissipative limit, $\eta \rightarrow 0$, Eq.~(23) recovers the 
dispersion relation derived by Bonanno \& Urpin (2006) for perturbations 
with $\vec{k} \cdot \vec{B} = 0$,
\begin{equation}
\sigma^{5} + \sigma^{3} (\omega^{2}_{0} + \kappa^{2})
+ \sigma^{2} \omega^{3}_{B \Omega} + \sigma \mu \kappa^{2} \omega^{2}_{0}
+ \mu \kappa^{2} \omega^{3}_{B \Omega} = 0. 
\end{equation}
The authors argued that Eq.~(25) has unstable solutions even when the 
criteria of the magnetorotational and Rayleigh instability are not satisfied. 
The condition of instability is $\omega_{B \Omega} \neq 0$ and, hence, the 
shear-driven instability arises if $B_{s} \neq 0$. The paper by Bonanno \& Urpin
(2007) considers the dispersion relation for any wavevectors, 
but does so in a high-conductivity limit $\eta \rightarrow 0$. 
 
Eq.~(23) can be simplified in many cases of interest if we take into account 
that, likely, the $s$-component of the magnetic field in disks is greater 
than the $z$-component, $B_s \gg B_z$. For perturbations with $\vec{k} \cdot 
\vec{B} = 0$, we have $k_s = - k_z B_z/ B_s$. Then, 
\begin{equation}
\mu = \frac{k_z^2}{k_z^2 + k_s^2} = \frac{B_s^2}{B_s^2 + B_z^2} \approx 1 . 
\end{equation}
Substituting $\mu \approx 1$ into Eq.~(23), we can transform it into
\begin{eqnarray}
(\sigma^2 + \kappa^2) [ \sigma^4 + 2 \omega_{\eta} \sigma^3 +
(\omega_{\eta}^2 + \omega_0^2) \sigma^2 +   \nonumber \\
(\omega_{\eta} \omega_s^2 + \omega_{\eta} \omega_0^2 + \omega_{B  \Omega}^3) 
\sigma
+ \omega_{\eta}^2 \omega_{s}^2] = 0.
\end{eqnarray}
Two roots of this equation describe the inertial waves which can be unstable
only if the Rayleigh criterion is satisfied. Other four modes are described 
by the dispersion relation 
\begin{eqnarray}
\sigma^4 + 2 \omega_{\eta} \sigma^3 + (\omega_{\eta}^2 + \omega_0^2) \sigma^2 
+  (\omega_{\eta} \omega_s^2 + \omega_{\eta} \omega_0^2 + \omega_{B \Omega}^3) 
\sigma
\nonumber \\
+ \omega_{\eta}^2 \omega_{s}^2 = 0.
\end{eqnarray}
This equation describes fast and slow magnetoacoustic waves, and we consider the
stability of these modes. Note that simplification (26) is made for
mathematical convenience rather than for physical relevance. In fact,  the 
dispersion relation (28) for fast and slow magnetosonic waves will change
little if $\mu \neq 1$. This particularly concerns the modes with $|\sigma| > \Omega$
because the dispersion equation for them will differ from Eq.~(28) only by
terms of the order of $(\mu -1) \kappa^2/ \sigma^2$ in coefficients. Therefore, our 
results can be applied with a sufficient accuracy also for the perturbations 
with $\mu \neq 1$.

\section{Criteria of instability in protostellar disks}

The Hurwitz theorem states that an equation of the fourth order,
\begin{equation}
\sigma^4 + b_3 \sigma^3 + b_2 \sigma^2 + b_1 \sigma^1 + b_0 =0,
\end{equation}
has at least one root with a positive real part (unstable mode) if one of 
the following inequalities
\begin{eqnarray}
\lefteqn{b_{3} < 0 \;, \;\;\; b_{0} <0 \;,}
\\
\lefteqn{b_{3} b_{2} - b_{1} < 0 \;, }
\\
\lefteqn{b_{1} (b_{3} b_{2} - b_{1}) - b_{3}^2 b_{0} <0 \;,}
\end{eqnarray}
is fulfilled (see Aleksandrov et al. 1985). Conditions (30) are never satisfied 
in protostellar disks. Therefore, the instability arises if one of the
inequalities (31) or (32) are satisfied. These inequalities can be rewritten 
for Eq.~(28) as
\begin{eqnarray}
\omega_{\eta} (2 \omega_{\eta}^2 + \omega_m^2) - \omega_{B \Omega}^3 < 0, \\
\omega_{B \Omega}^6 \! - \! 2 \omega_{B \Omega}^3 \omega_{\eta} 
(\omega_{\eta}^2 \! - \! \omega_s^2) \! - \! \omega_{\eta}^2 \omega_m^2 
( \omega_0^2 + 2 \omega_{\eta}^2 + \omega_s^2 ) \! > \! 0.
\end{eqnarray}
Generally, the expressions on the l.h.s. of both these inequalities can 
have a positive or negative sign, depending on the value of $\omega_{B \Omega}$.
Therefore, Eq.~(33) and (34) impose restrictions on the rate of differential 
rotation which should be greater than some critical value.
Note that, if the azimuthal field is generated by winding up the radial field,
then $B_s B_{\varphi} \Omega' >0$ and, hence, $\omega_{B \Omega}^3 > 0$ as 
well. Therefore, we consider only this case. Denoting $\delta = c_{As} 
c_{A \varphi}/c_{A}^2$, we have from Eq.~(33)
\begin{equation}
\delta \frac{H s \Omega'}{c_A} > \frac{\eta}{c_A H} \left[ 
\frac{2 \eta^2}{c_A^2 H^2} (kH)^4 + (kH)^2 \right].
\end{equation} 
This condition can be satisfied only if the differential rotation is sufficiently 
strong. Since $\Omega' < 0$ in astrophysical disks, we consider only this 
case. Then, we have $\delta < 0$ because $B_s B_{\varphi} \Omega' \propto \delta \Omega' > 0$. 
The r.h.s. of Eq.~(35) is a monotonously increasing function of $kH$ 
and is a minimum  for the minimal possible value of $kH$. Since a 
short-wavelength approximation applies only if $kH > 1$, we can estimate the 
domain of instability supposing $kH=1$ in inequality (35). Then, in order to 
fulfill the instability condition (35) for short-wavelength perturbations 
with $kH >1$, the rotation shear has to satisfy, at least, the inequality:
\begin{equation}
\delta \frac{H s \Omega'}{c_A} > \frac{\eta}{c_A H} \left( 
\frac{2 \eta^2}{c_A^2 H^2} + 1 \right), 
\end{equation}
In Fig.~1, we plot the regions of instability given by Eq.~(36) for
different negative values of $\delta$. At given $\delta$, the instability can 
occur if the value of $|H s \Omega' /c_A|$ is above the corresponding curve 
(note that $\Omega'$ is negative in Fig.~1). If the magnetic and thermal 
energies are comparable and departures from the Keplerian disk are not large, 
then $c_A/H \sim c_s/H \sim \Omega$, and $H s \Omega'/c_s \sim s \Omega' /
\Omega \approx 3/2$. Inequality (36) can be satisfied only for relatively 
small values of $\eta / H c_A < |\delta|$, or $|\delta| \sim B_s/B_{\varphi} 
> 1/ {\rm Re}_m$. { This condition can be satisfied only if the Hall
parameter is large $a_{\rm e} > 1$, because $\delta \sim 1/{\rm Re}_m$ (see 
Eq.~(13)) in the case $a_{\rm e}<1$. Condition (36) can also be fulfilled 
in a non-Keplerian disk.} For a non-Keplerian disk, the quantity $H s \Omega'
/ c_{A}$ can vary in a wider range even if the magnetic and thermal energies 
are comparable, but condition (36) still cannot be satisfied for relatively 
large values of $\eta/ H c_{A}$.    

\begin{figure}
\begin{center}
\includegraphics[width=9.0cm]{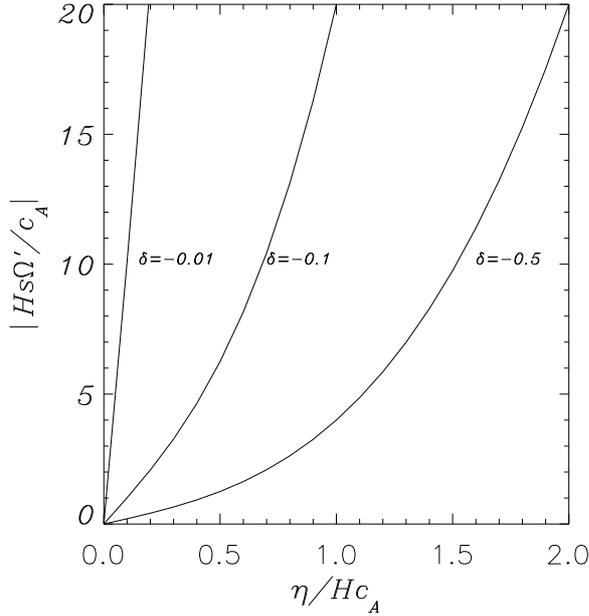}
\caption{The region of parameters where the instability can occur in 
accordance with Eq.~(32) for different values $\delta$.}
\end{center}
\end{figure}

By analogy, we can also transform  condition (34). The minimal unstable
rotation shear corresponds again to the maximal possible wavelength. Again,
the maximum wavelength can be estimated from the condition $kH \sim 1$. For 
such wavevectors, Eq.~(36) can be transformed
into:
\begin{eqnarray}
\delta^2 \left( \frac{H s \Omega'}{c_A} \right)^2 > 2 \delta \left(
\frac{H s \Omega'}{c_A} \right) \frac{\eta}{H c_A} \left( 
\frac{\eta^2}{H^2 c_A^2} - \beta \right) + \nonumber \\
\frac{\eta^2}{H^2 c_A^2} \left( 1 + 2 \beta + 
\frac{2 \eta^2}{H^2 c_A^2} \right), 
\end{eqnarray}  
where $\beta = c_{s}^2/c_{A}^2$. 

\begin{figure}
\begin{center}
\includegraphics[width=9.0cm]{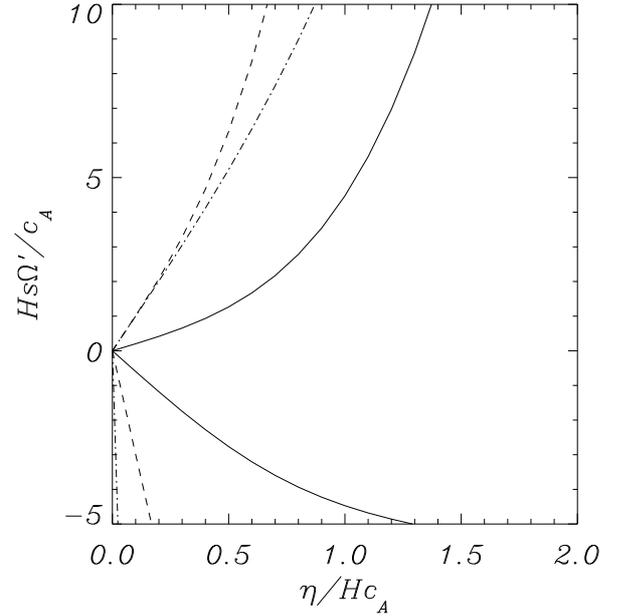}
\caption{The region of parameters where instability can occur in
accordance with Eq.~(33). The values of $[\delta, \beta]$ are [-0.5, 1] (solid 
lines), [-0.1, 1] (dashed lines), and [-0.1, 10] (dashed-and-dotted lines).}
\end{center}
\end{figure}

In Fig.~2, we show the region of parameters where the instability can occur
in accordance with Eq.~(37). Inequality (37) is quadratic in
$H s \Omega'/c_A$ and can be satisfied if this parameter is larger/smaller
than the largest/smallest root of the corresponding quadratic equation. 
Therefore, for a given combination of $\beta$ and $\delta$, the regions of instability lie
above the upper line and below the lower line.

\section{Growth rate of instability}

To calculate the instability growth rate, it is convenient to 
introduce dimensionless quantities,
\begin{eqnarray}
\Gamma= \frac{\sigma}{\Omega} \;,\;\; 
\alpha = \frac{\eta}{\Omega H^2} \;,\;\;
\varepsilon = \frac{c_{s}}{H \Omega} \;,\;\;
x = kH\;,
\nonumber \\
\delta = \frac{c_{A \phi} c_{A s}}{c^{2}_{A}} \;,\;\;
q = \left| \frac{s \Omega'}{\Omega} \right| \;,
\nonumber 
\end{eqnarray}
(we assume $B_s B_{\varphi} \Omega' >0$). Then, Eq.~(24) transforms into
\begin{eqnarray}
\Gamma^{4} + 2 \alpha x^2 \Gamma^{3} + 
+ \Gamma^{2} x^2 [ \alpha^2 x^2 + \varepsilon^2 (1 + 1/\beta)] +
\nonumber \\
+ \Gamma \varepsilon^2 x^2 [ \alpha x^{2} (2 + 1/\beta) + q \delta /\beta ]
+ \alpha^2 \varepsilon^2 x^{6} =0.
\end{eqnarray}
The dependence on the wavelength is characterized by the parameter $x^{2}$
in this equation. The parameter $\alpha$ is of the order of $1/{\rm Re}_m$. 
We solved eq.~(38) numerically for different values of 
the parameters by computing the eigenvalues of the matrix whose 
characteristic polynomial is given by Eq.~(38) (see Press et al. 1992 for 
details).

\begin{figure}
\begin{center}
\includegraphics[width=9.0cm]{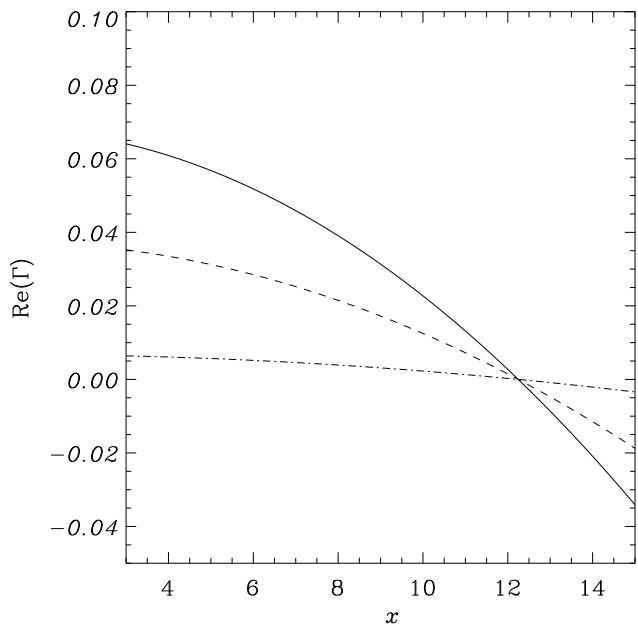}
\caption{The dependence of the growth rate on $x=kH$ for the Keplerian
disk, with $q=3/2$ and $\varepsilon=1$, and for $\beta=0.1$ (solid line), 
$1$ (dashed line), and $10$ (dashed-and-dotted line). The other parameters
are $\delta = 0.1$ and $\alpha = 0.001$.}
\end{center}
\end{figure}

In Fig.~3, we plot the dependence of the growth rate on $x$ for the
Keplerian disk with $q=3/2$ and $\varepsilon =1$ and for several values of the
parameter $\beta$. In all cases, the growth rate decreases monotonically with 
decreasing  wavelength $\lambda = 2 \pi/k$ because ohmic dissipation is 
more efficient for perturbations with a shorter wavelength.  
For any ratio of the magnetic and gas pressures, the instability does not 
occur if $kH > 12$. If the magnetic pressure is greater than the gas 
pressure ($\beta < 1$), the growth rate can reach relatively large values 
$\Gamma \approx 0.04-0.06$. This corresponds to the growth time only $3-4$ 
times longer than the rotation period. The magnetorotational instability 
cannot arise in such a strong magnetic field no matter how large is the 
electrical conductivity of gas. Even if $\vec{k}$ is not perpendicular 
to $\vec{B}$, the MRI can occur only if the magnetic field and 
wavevector satisfy the condition $k^2 c_A^2 < 2 s \Omega \Omega'$. Assuming 
that $s \Omega' \sim \Omega$, we can transform this inequality approximately 
into $\Omega > c_A k \sim x c_s/H \sqrt{\beta}$. Since $c_s/H \sim \Omega$ in 
the Keplerian disk, we find that MRI arises if $x /\sqrt{\beta} <1$. In the 
case $\beta < 1$, this condition never applies and the MRI is suppressed. 
On the contrary, the shear-driven instability turns out to be rather efficient
in such a strong field.

\begin{figure}
\begin{center}
\includegraphics[width=9.0cm]{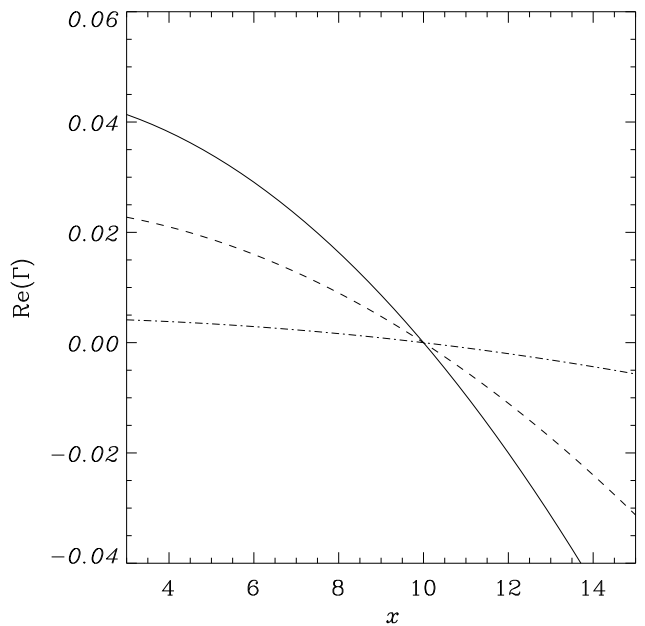}
\caption{The same as in Fig.~3 but for a nonKeplerian disk with $q=1$ and
$\varepsilon=3$. }
\end{center}
\end{figure}

In Fig.~4, we plot the same dependence, but for the nonKeplerian disk with
$q=1$ and $\varepsilon =3$. The other parameters are same. Qualitatively,
the dependences are very similar to those shown in Fig.~3. The instability
grows faster in the disk with larger $\beta$, which seems to be a general 
feature of the shear-driven instability. However, the growth rate is typically 
smaller for a non-Keplerian disk. The range of unstable wavevectors is 
limited approximately by $kH < 10$. Note that the half-thickness $H$ is 
essentially larger in a non-Keplerian disk than in the Keplerian one. 

\begin{figure}
\begin{center}
\includegraphics[width=9.0cm]{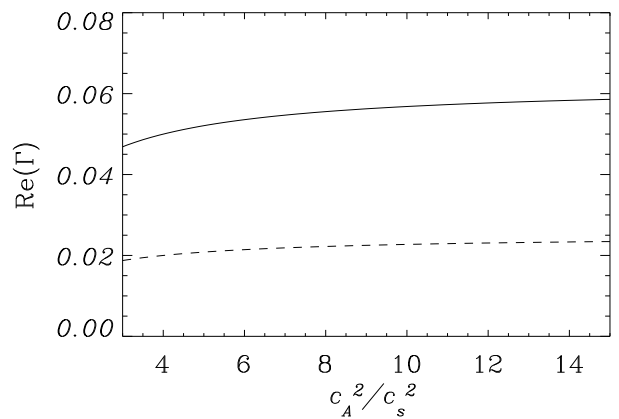}
\caption{The dependence of the growth rate on $\beta$ for the Keplerian disk
with $q=3/2$, $\varepsilon=1$, $\delta =0.1$, and $\alpha=0.001$. The solid 
and dashed lines correspond to $x=5$ and $10$, respectively.}
\end{center}
\end{figure}

The dependence of the growth rate on $\beta^{-1}= c_{A}^2 / c_{s}^2$ is shown in Fig.~5 
for the 
disk with $q=3/2$ and $\varepsilon = 1$. At a given wavelength,
the growth rate increases with $1/\beta$ slowly approaching a saturation
value that depends on $q$. The saturation growth rate is $\sim 0.06 \Omega$
and $0.025 \Omega$ for $q=5$ and $10$, respectively. This correspond to the 
growth time of about $\sim3$ and $\sim6$ rotation periods. The dependence 
of the shear-driven instability on the magnetic field is unusual, 
at least, in the domain of parameters because usually a strong 
magnetic field tends to suppress other types of instability.
\begin{figure}
\begin{center}
\includegraphics[width=9.0cm]{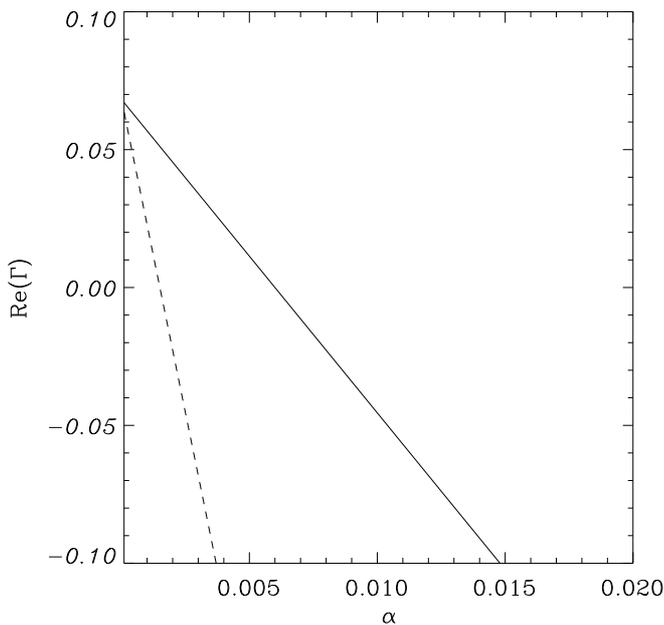}
\caption{The dependence of the growth rate on $\alpha$ for $q=3/2$, 
$\varepsilon=1$, $\delta=0.1$, and $\beta=0.1$. The solid and dashed curves
correspond to $q=5$ and $10$, respectively.}
\end{center}
\end{figure}
Fig.~6 shows the dependence of the growth rate on the parameter $\alpha$, which
is approximately equal to the inverse magnetic Reynolds 
number, $\alpha \sim 1/{\rm Re}_m$. Therefore, higher values of $\alpha$ 
correspond to a higher dissipation rate, and the instability should be 
suppressed at large $\alpha$. Indeed, the growth rate goes to 0 if $\alpha 
\sim 0.001-0.01$, depending on the wavelength. Hence, one can expect that the
shear-driven instability occurs even in disks with a relatively small 
magnetic Reynolds number $\sim 100$.

\section{Discussion}
We have considered the new instability that can arise in protostellar 
disks under the combined influence of the magnetic field, differential 
rotation, and compressibility. To illustrate the main qualitative features 
of the instability, we analyzed a particular case of perturbations with the 
wavevector $\vec{k}$ perpendicular to the magnetic field $\vec{B}$. In this 
case, the standard MRI does not occur because its growth rate is $\propto 
(\vec{k} \cdot \vec{B})$. The considered instability is related 
to shear and compressibility of a magnetized gas. In the incompressible 
limit ($c_{s} \rightarrow \infty$) we have, from Eq.~(27),
\begin{equation}
(\sigma^2 + \kappa^2) (\sigma + \omega_{\eta})^{2} =0,
\end{equation}
and the instability does not occur. This is a major difference from other 
instabilities caused by differential rotation, such as the Rayleigh or 
magnetorotational instabilities that can occur in incompressible limit. 

The magnetic field is likely thermal or superthermal in protostellar disks 
(see Donati et al. 2005). The considered instability can arise even in a 
very strong magnetic field when the MRI is suppressed completely, It is 
known that the MRI does not occur if the magnetic field satisfies the 
condition $B > B_{cr}= (s \Omega' /k) \sqrt{4 \pi \rho}$. The shear-driven 
instability can operate even if the field is stronger than $B_{cr}$. The 
growth time of instability is still rather short (several rotation periods) 
in the gas where the magnetic pressure is greater than the thermal pressure. 

The criteria for the considered instability can be satisfied in 
protostellar disks. Likely, all disks have both radial and azimuthal magnetic 
fields. These magnetic fields are found in observations, as well as in 
numerical simulations. Superthermal magnetic fields do not suppress the
instability  crucial in cold protostellar disks.  The instability can arise even in 
a low conductive gas where the magnetic Reynolds number associated with 
rotation is relatively small, Re$_m \sim 100$.       

Our paper considers only a local axisymmetric instability. Most likely, 
the same mechanism of instability will efficiently destabilize 
perturbations with the wavelength comparable to the disk scale height. 
Dissipative effects are less important for these perturbations, and the unstable modes 
can grow even faster. A global instability caused by a combined influence 
of differential rotation and compressibility will be considered elsewhere.
The turbulence that could be generated by the considered instability can be 
strongly anisotropic in the $(s,z)$-plane because both the criteria and 
growth rate are sensitive to the direction of the wave vector. The 
generated turbulence, however,  might be efficient in the radial transport of angular 
momentum in strongly magnetized protostellar disks.

\section*{Acknowledgments}
This research project has been supported by a Marie Curie Transfer of
Knowledge Fellowship of the European Community's Sixth Framework
Programme under contract number MTKD-CT-002995.
VU thanks also INAF-Osservatorio Astrofisico di Catania for hospitality.

{}

\end{document}